# Prediction of Political Leanings of Chinese Speaking Twitter Users

Fenglei Gu∗
Department of Computer Science
New York University
fenglei.gu@nyu.edu

Duoji Jiang∗
Department of Computer Science
New York University
duoji.jiang@nyu.edu

∗Both authors contributed equally to this paper

*Abstract*—This work presents a supervised method for generating a classifier model of the stances held by Chinese-speaking politicians and other Twitter users. Many previous works of political tweets prediction exist on English tweets, but to the best of our knowledge, this is the first work that builds prediction model on Chinese political tweets. It firstly collects data by scraping tweets of famous political figure and their related users. It secondly defines the political spectrum in two groups: the group that shows approvals to the Chinese Communist Party and the group that does not. Since there are not space between words in Chinese to identify the independent words, it then completes segmentation and vectorization by Jieba, a Chinese segmentation tool. Finally, it trains the data collected from political tweets and produce a classification model with high accuracy for understanding users' political stances from their tweets on Twitter.

## I. Introduction

### A. Background and problem description

Social media is used by people to express their opinions and views. A large part of the population shares news of politics they support. In addition, some of people show strong political leanings on their user profiles. Therefore it is possible to distinguish people who support one political party from another on social media platforms. Among social media platforms, Twitter is a good candidate to be tested on. Because of the limited length of tweets, users tend to craft their statements to express their political stances. Tweets have less unrelated data, and the accuracy is higher than other platforms that do not have word limits.

This paper presents a supervised method for understanding the stances held by politicians and other Twitter users who speak Chinese. The political stance is represented by a political spectrum. A political spectrum is a system to classify and to differentiate political positions in some geometric axes that represent independent political dimensions. We define the political spectrum as two groups: the group that shows approvals to the Chinese Communist Party and the group that does not. We will carefully select users who are Chinese speakers and create a pro-con model.

### B. significance

We select users who are Chinese speakers and create a pro-con model for the following reasons:

- To the best of our knowledge, it is the first time to do research on political leanings of Chinese speakers. The mainstream is to research on English speakers, while fewer researchers choose to analyze minorities. This paper can provide a different view on patterns of minorities.

- Unlike most Americans, fewer Chinese speakers show their approval/disapproval on the ruling party in the U.S. Instead, they express their opinions on Chinese Communist Party. Therefore it is possible to create a special pattern for understanding these users' political stances.

- Because the Chinese speaker community on Twitter is relatively small compared to the English speaker community and most Chinese speakers do not use Twitter as their social media platform, it is very unlikely to see topics in Chinese speaker community trending on Twitter. We can find what is more important to this minor community and the different language patterns of both sides by collecting data from their tweets.

- With the data we collect, we may define terms in Chinese that include illegal contents, such as extreme violence, drug trade, or prostitution. Thereby helping detect users who violate Twitter rules.

Our algorithm will train the models using data collected from political Tweets in Chinese and generate a model for deciding users' opinions on Chinese Communist Party. The algorithm will be specified in the following sections.

## II. Related Literatures

Bakliwal et al employed supervised learning based on annotated sentiment of political tweets. They performed on a dictionary of sentiment score for each word and subjective lexicon, achieved an accuracy of 61.62\%. [1]

Kristen Johnson and Dan Goldwasser primarily used the probability soft logic in combination with several weakly-supervised local models, based on similarities and differences of politician's statement on certain frames.[2]

Daniel Preotiuc-Pietro, Ye Liu, Daniel Hopkins, and Lyle Ungar performed a political stance prediction on a seven-point scale, to predict the political stances of less-active users.[3]

Kristen Johnson, I-Ta Lee, and Dan Goldwasser presented a joint model that combines both linguistic features and ideological phrase, extracted by a state-of-the-art embedding model of tweets to predict the general frame of political tweets.[4]

Peter Stefanov, Kareem Darwish, Atanas Atanasov, and Preslav Nakov presented an unsupervised approach of predicting political stance based on overall twitter behaviors including retweeting and the simultaneity of tweeting.[5]

## III. METHODOLOGY

T-Scrapper, an open-source Twitter Tweet data crawling tool that overrides some limits of the official Twitter API, is used to collect the training data - tweets (text documents) sent by Chinese Twitter users. Then, Jieba, together with Open Chinese Convert (OpenCC), is used to segment these documents of Chinese texts into vectors of words, which involves Hidden Markov Model (HMM) and Viterbi Algorithm. After that, K Nearest Neighbor (k-NN) is used for classification, where the distance between documents is determined by Term Frequency and Inverse Document Frequency (TF-IDF) and Cosine Similarity. Finally, cross validation is used to make full use of the limited training data, and to ensure the quality of the trained model.

### A. Jieba

Jieba is a Chinese text segmentation module. Unlike English, there is no white space between words in Chinese, thus one would need to figure out what the words are in the given texts. This is exactly why Jieba can help. It parses Chinese texts into tokens, combining single Chinese characters into meaningful words. In the case of rare or new words that are not included in in Jieba's default built-in dictionary, an "HMM-based model is used with the Viterbi algorithm"; or users may add new words to build their own dictionary.[6]

Jieba is claimed to possess the capability of parsing Chinese characters in both simplified form and traditional form. However, based on our experiment, it is found that its capacitance is limited for texts in traditional Chinese. Thus, "Open Chinese Convert (OpenCC)" is introduced below to translate traditional Chinese characters in to simplified form.

### B. Viterbi Algorithm

The Viterbi algorithm is a dynamic programming algorithm that produces a sequence of hidden states that has the most probability. In the context of HMM, the Viterbi algorithm gives the most likely POS tag sequence for tokens. For every token, it calculates the transition probability and emission probability, and find the combination of tags that has the highest probability.

### C. Open Chinese Convert(OpenCC)

Open Chinese Convert (OpenCC) is an open-source tool that can be used to convert Traditional Chinese characters into Simplified Chinese. It is used to improve the accuracy of segmentation by Jieba.

### D. Term Frequency and Inverse Document Frequency(TF-IDF)

Term Frequency and Inverse Document Frequency (TF-IDF) is a numerical statistic that count the importance of a word to a document in a corpus. Term Frequency (TF) of term (t), denoted as $T(t)$, is the number of times the term occurs in the document (or the number of times divided by the length of the document). Inverse Document Frequency (IDF) of term (t), denoted as $I(t)$, is the number of documents in the corpus C, divided by the number of documents $d_i$ that contains the term, normalized with log function. TF-IDF $TI(t)$ is the product of TF and IDF.

$$I(T) = \log\left(\frac{|C|}{d_i \in C, t \in d_i}\right) \quad (1)$$

$$TI(t) = T(t) \cdot I(t) \quad (2)$$

### E. Cosine Similarity

Cosine similarity is Cosine of the angle between the Vectors. It defines the similarity between documents. The value is higher if words in two documents are more similar. The value is 1 if two documents are identical and 0 if two documents don't contain common words.

$$Similarity(A, B) = \frac{\sum_i a_i \times b_i}{\sqrt{\sum_i a_i \times \sum_i b_i}} \quad (3)$$

### F. K-Nearest-Neighbor (K-NN)

K-nearest-neighbors algorithm (k-NN) is a non-parametric classification method used for classification and regression. In both cases, the input consists of the k closest training examples in data set. The output depends on whether k-NN is used for classification or regression. In this paper, k-NN is used for classification.

In k-NN classification, the output is a class label. An object is classified by k votes of its nearest neighbors. If k = 1, then the object is assigned to the class of that single nearest neighbor. k-NN is a type of classification that the function is only approximated locally. It can improve the accuracy by normalizing the data if the features come in very different scales. A useful technique to normalize the data is to assign weights to the contributions of the neighbors, which is to assign more weights to the nearer neighbors than the other ones.

### G. Cross Validation

Cross-validation is a statistic model that evaluates the skill of machine learning models. There are two types of cross-validation: Exhaustive cross-validation and non-exhaustive cross-validation. Exhaustive cross-validation methods learn and test on all possible ways to divide the original sample into a training and a validation set, and non-exhaustive cross-validation methods do not. In this paper, we will use K-fold cross-validation. We will divide the data into k data sets and choose 1 set as the testing set and the others as training sets. And we repeat the process for k times. This will result in a less biased evaluation of our machine learning model.

## IV. DATA SET

### A. Obtain Data

We obtain data primarily based on political leanings of Twitter users. We divide political leanings into the group that shows approvals to the Chinese Communist Party and the group that does not. The process of finding Twitter accounts is started by arbitrarily picking some accounts whose political leaning is known. For example, we first pick New China Press (@XinhuaChinese), a Chinese state-affiliate media, which has a well-known pro-Beijing leaning. Then, we search in "You Might Like", in the comment area, and from retweeted tweets, to see if there is any related account that has the same political leaning. Similarly, we pick accounts that are known as pro-Democracy, such as Amnesty International (@amnestychinese), and search for related accounts. We set 10,000 followers as the baseline and discard accounts that have fewer than 10,000 followers. Also discarded are inactive accounts that have sent less than 10 tweets in 2021, as of April 15.

### B. Label Data

We define political leanings of those do not have a well-known leaning by their views of some global affairs and word choices. For example, we manually label users who support China on human rights issue as "pro-Beijing" and those who condemn as "anti-Beijing". We obtain 40 pro-Beijing accounts and 41 pro-Democracy accounts, 81 in total. We set the time interval to be from January 1, 2021, to April 15, 2020. And then we scrap these accounts to get their tweets in the text form between this time interval. We use Open Chinese Convert to convert traditional Chinese characters into simplified Chinese characters and use Jieba for text segmentation. This process produces one text document for each account that contains the result of text segmentation.

### C. Split Data

These 81 accounts above are further split into test set (21 accounts, including 10 pro-Beijing and 11 pro-Democracy) and non-test set (60 accounts, including 30 pro-Beijing and 30 pro-Democracy). The test set are only used in the final stage - testing; while the non-test set are iteratively split into training sets and validation sets, following the process of cross validation.

## V. TEST RESULTS

There mainly two types of measures in our evaluation process: accuracy and F1 score. Their definitions can be found in Appendix A.

### A. Baseline 0

Table 1: Confusion Matrix for Baseline 0

| Output / Key | Beijing | Democracy |
|---|---|---|
| Beijing | 0 | 10 |
| Democracy | 0 | 11 |

The accuracy is 0.52. The F1 score is 0 if "Beijing" is treated as the positive category; or 0.69 if "Democracy" is treated as the positive category.

### B. Baseline 1

A more complex system is one that uses 5-Nearest Neighbor where distance between two accounts is defined as the number of words that appeared only in one account's list of of top 25 most frequently used words in their tweets. The result is as below (See Table 2)

Table 2: Confusion Matrix for Baseline 1

| Output / Key | Beijing | Democracy |
|---|---|---|
| Beijing | 10 | 0 |
| Democracy | 5 | 6 |

The accuracy is 0.76. The F1 score is 0.80 if "Beijing" is treated as the positive category; or 0.71 if "Democracy" is treated as the positive category. It is unknown why the system tends to classify data as pro-Beijing.

### C. Our Model based on k-NN with TF-IDF

Our Model is based on 5-Nearest Neighbors to make predictions of the test set's political labels with TF-IDF to define the distance between tweeter accounts. The labels of tweeter accounts of the Test set are hidden from the validation set, and their distance with each of the tweeter accounts of the non-test set calculated by cosine similarity of TF-IDF is recorded and ranked. 5 Nearest Neighbors' political labels are used in a voting, the prediction is that the majority of the labels among the 5 accounts' labels predict the political label of the test set.

Table 3: Confusion Matrix for our Model

| Output / Key | Beijing | Democracy |
|---|---|---|
| Beijing | 9 | 10 |
| Democracy | 0 | 11 |

The accuracy is 0.95. The F1 score is 0.95 if "Beijing" is treated as the positive category; or 0.96 if "Democracy" is treated as the positive category.

## VI. DISCUSSION

### A. Analysis of Strength and Potential Issues

The advantages of our approach are as follows. Firstly, the approach is: simple - only k-NN, TF-IDF, cosine similarity is applied with no other complex algorithms; fast - Not including the time of scrapping tweets, the time cost of our model analyzing 21 accounts is only 3 minutes; high accuracy - this simple approach achieved a high accuracy of 0.95.

When analyzing the one twitter account with which the model makes wrong classification, it is noted that it is labeled as pro-Beijing, yet it is not an identical pro-Beijing twitter account, although with an identifiable pro-Beijing political leaning, majority of its posts focus on the American politics. The potential issue could be that the sample is small and may not be inclusive enough, as there can be multiple subdivisions in one political leaning that are not included.

Besides, our labels are manually done by researchers without a verification of third parties. The accuracy would also be negatively influenced if one Twitter account often tweets things unrelated to their political leanings, such as their daily life. Finally, as our results are directly related to the dictionary used in Jieba segmentation tool, the accuracy would also be negatively impacted by the inaccuracy of the segmentation.

Finally, the size of the data set is not large enough, and some accounts are owned by the same person or organization.

### B. Alternative Solutions

There are many other Machine Learning techniques to solve classification problems, such as Support Vector Classification, Logistic Regression, Decision Trees, etc.

### C. Future Improvements and Recommendations

One of the most critical improvement would be to increase the volume of sample (training data).

Besides, the k-NN method could be modified. In our model, the decision of classification is based on unweighted vote of the nearest 5 neighbors. However, it is also possible to realize unweighted voting where the nearer neighbors get more weights than relatively further neighbors.

Furthermore, the classification process could involve multiple (non-binary) classes. Currently, all the accounts are only labelled as either pro-Beijing or pro-democracy. However, it would be more inclusive and closer to reality if there are more classifications of labels.

Last but not least, our research could potentially be extended into an unsupervised learning process with k Means, thereby revealing likely emerging political trends, or discovering possibly other potential results that reveals certain patterns.

## VII. CONCLUSION

We have presented a supervised method for predicting the political leaning of Twitter users. The political leanings defined in the paper are whether users approve or disapprove of the Chinese Communist Party. Our method takes data from Twitter users who speak Chinese. We scrap these users' Twitter accounts that have more than 10,000 followers to get tweets in the text form and use tools for text segmentation. We use K-Neighbor method with TF-IDF and Cosine similarity to produce a pattern for classifying users' political leanings. This method achieves a high F1 accuracy of 0.95. Our research reveals that it is feasible to classify active users' political leanings based on their tweets. But it has some potential issues, such as the insufficient sample volume and inaccurate tagging. Future improvement is needed, including increasing the sample volume and modifying the algorithm.

ACKNOWLEDGMENT

We would like to express our special thanks to Prof. Adam Meyers at Department of Computer Science, New York University, for his great efforts in teaching us the crucial ideas and knowledge in his Natural Language Processing course.

We would also like to deliver our sincere gratitude to Sun Junyi (Github User: fxsjy) and other contributors, for their wonderful open-source project, Jieba, which helped us segmenting training texts; and Carbo Kuo (Github User: BYVoid) and other community members, for their amazing open-source project, OpenCC, which helped us to convert traditional Chinese characters into their simplified counterparts. You may find them online at https://github.com/fxsjy/jieba and https://github.com/BYVoid/OpenCC, respectively.

We would also thank our friend, Xinyuan Zhao, at department of Computer Science, New York University, who provided a lot of great advises and help on writing the paper.

Last but not least, we would like to convey our gratefulness to our parents, who have long been spiritually and financially supporting us in many ways. Without them, this paper could not have been finished and presented here today.

## APPENDEX

### A. Concept Explanations

For each category of political leanings, we define 4labels, based on the system output and the true answer key: True Positive (TP), True Negative (TN), False Positive (FP), False Negative (FN). (See Table 4)

Table 4: Confusion Matrix for Republican Party

| Output / Key | GOP | Non-GOP |
|---|---|---|
| GOP | TF | FN |
| Non-GOP | FP | TN |

We further define" Accuracy"," Precision"," Recall" and" F1- Score" as below

$$Accuracy = \frac{\#TP + \#TN}{\#Total} \quad (4)$$

$$Precision = \frac{\#TP}{\#TP + \#FP} \quad (5)$$

$$recall = \frac{\#TP}{\#TP + \#FN} \quad (6)$$

$$F1 = \frac{2 \cdot Precision \cdot Recall}{Precision + Recall} \quad (7)$$

## B. Source Codes and Corpora Data

All the codes and data can be found on our website, https://cs.nyu.edu/ fg1121/twitter-politics.